# Prediction of spin orientations in terms of HOMO-LUMO interactions using spin-orbit coupling as perturbation

Myung-Hwan Whangbo[a,*], Elijah E. Gordon[a], Hongjun Xiang[b], Hyun-Joo Koo[c], and Changhoon Lee[d]

[a] Department of Chemistry, North Carolina State University, Raleigh, NC 27695-8204, USA

[b] Key Laboratory of Computational Physical Sciences (Ministry of Education), State Key Laboratory of Surface Physics, Collaborative Innovation Center of Advanced Microstructures, and Department of Physics, Fudan University, Shanghai 200433, P. R. China

[c] Department of Chemistry and Research Institute for Basic Science, Kyung Hee University, Seoul 130-701, Republic of Korea

[d] Department of Chemistry, Pohang University of Science and Technology, Pohang 790-784, Korea

**Conspectus**

For most chemists and physicists, electron spin is merely a means needed to satisfy the Pauli principle in electronic structure description. However, the absolute orientations of spins in coordinate space can be crucial in understanding the magnetic properties of materials with unpaired electrons. At a low temperature the spins of a magnetic solid may undergo a long-range magnetic ordering, which allows one to determine the directions and magnitudes of spin moments by neutron diffraction refinements. The preferred spin orientation of a magnetic ion can be predicted on the basis of density functional theory (DFT) calculations including electron correlation and spin-orbit coupling (SOC). However, most chemists and physicists are unaware of how the observed and/or calculated spin orientations are related to the local electronic structures of the magnetic ions. This is true even for most crystallographers who determine the directions and magnitudes of spin moments because, for them, they are merely the parameters needed for the diffraction refinements. The objective of this article is to provide a conceptual framework of thinking about and predicting the preferred spin orientation of a magnetic ion by examining the relationship between the spin orientation and the local electronic structure of the ion. In general, a magnetic ion *M* (i.e., an ion possessing unpaired spins) in a solid or a molecule is surrounded with main-group ligand atoms *L* to form an $ML_n$ polyhedron, where n is typically 2 – 6, and the *d*-states of $ML_n$ are split because the antibonding interactions of the metal *d*-orbitals with the *p* orbitals of the surrounding ligands *L* depend on the symmetries of the orbitals involved.[1] The magnetic ion *M* of $ML_n$ has a certain preferred spin direction because its split *d*-states interact among themselves under SOC.[2,3] The preferred spin direction can be readily predicted on the basis of perturbation theory, in which the SOC is taken as perturbation and the split *d*-states as unperturbed states, by inspecting the magnetic quantum numbers of its *d*-orbitals present in the HOMO and LUMO of the $ML_n$ polyhedron. This is quite analogous to how chemists predict the allowedness/forbiddenness of a chemical reaction in terms of the HOMO-LUMO interactions by simply inspecting the symmetries of these frontier orbitals.[4,5] Experimentally, the determination of the preferred spin orientations of magnetic ions requires a sophisticated level of experiments, for example, neutron diffraction measurements for magnetic solids with ordered spin state at a very low temperature. Theoretically, it requires an elaborate level of electronic structure calculations, namely, DFT calculations including electron correlation and SOC. We show that the outcomes of such intricate experimental measurements and theoretical calculations can be predicted by a simple perturbation theory analysis.

**Introduction**

An important role of a theory is to provide accurate predictions that can be tested by experiments. At the current level of DFT calculations including electron correlation and SOC, the preferred spin orientations of magnetic ions in a given magnetic solid can be easily determined by calculating its total energy as a function of their spin orientation. The results of such calculations are mostly in agreement with the spin orientations observed at low temperatures by neutron diffraction refinements. However, these calculations and experiments present no information about how the calculated and observed spin orientations are related to the local environments of the magnetic ions. To answer this question, one needs a qualitative theory that provides a conceptual picture with which to organize and think about experimental observations [6] and hence enables us to anticipate the outcome of new experiments or calculations to perform. The *d*-states of a transition-metal magnetic ion *M* are split in energy because the strengths of the metal-ligand σ- and π-antibonding interactions depend on the site-symmetry of $ML_n$ and on the orbitals involved.[1] When SOC is introduced at the metal *M* (or at the ligands *L*), the split *d*-states interact among themselves because they no longer remain the eigenstates of $ML_n$ under the perturbation of the SOC. In this article we show that the preferred spin orientation of *M* can be easily predicted by analyzing the angular properties of the metal *d*-orbitals (or ligand *p*-orbitals) present in the HOMO and LUMO of $ML_n$.

**Angular properties of orbital and spin**

The magnetic orbitals (i.e., singly-filled *d*-states) of an $ML_n$ polyhedron are readily identified, once the split *d*-states and the spin S of the ion *M* are known. The *d*-orbitals of *M* present in the split *d*-states have the angular behaviors of $3z^2–r^2$, xz, yz, xy or $x^2$-$y^2$. (The $3z^2–r^2$ orbital is often referred to as the $z^2$ orbital.) To discuss how these angular behaviors are affected by SOC, we recall that they are linear combinations of orbital states (or spherical harmonics), $|L, L_z\rangle$, where L = 2, and $L_z$ = −2, −1, 0, +1, +2 for the *d*-orbitals. In terms of the magnetic quantum numbers $L_z$, the *d*-orbitals are grouped into three sets:

$$\begin{aligned} L_z &= 0 \quad \text{for } 3z^2 - r^2 \\ L_z &= \pm 1 \quad \text{for } \{xz, yz\} \\ L_z &= \pm 2 \quad \text{for } \{xy, x^2 - y^2\} \end{aligned}$$

Thus, as depicted in **Fig. 1a**, the minimum difference $|\Delta L_z|$ in the magnetic quantum numbers is 1 between $\{3z^2–r^2\}$ and {xz, yz} as well as between {xz, yz} and $\{xy, x^2–y^2\}$. $|\Delta L_z|$ = 2 between $\{3z^2–r^2\}$ and $\{xy, x^2–y^2\}$, while $|\Delta L_z|$ = 0 between xz and yz as well as between xy and $x^2–y^2$. In a similar manner, the *p*-orbitals are expressed as linear combinations of the orbital states $|L, L_z\rangle$, where L =1, and $L_z$ = 1, 0, -1. Thus, the *p*-orbitals are grouped into two sets:

$$\begin{aligned} L_z &= 0 \quad \text{for } z \\ L_z &= \pm 1 \quad \text{for } \{x, y\} \end{aligned}$$

As depicted in **Fig. 1b**, $|\Delta L_z|$ = 1 between z and {x, y}, and $|\Delta L_z|$ = 0 between x and y.

Given the SOC constant λ, the orbital angular momentum $\vec{L}$ and the spin angular momentum $\vec{S}$ of the metal ion *M*, the SOC term is written as $\lambda\vec{S}\cdot\vec{L}$ in classical mechanics. In quantum mechanics it is expressed as the operator $\lambda\hat{S}\cdot\hat{L}$, where $\hat{S}$ and $\hat{L}$ are the spin and orbital angular momentum operators, respectively. (In atomic units in which $\hbar$ = 1, $|\vec{L}| = L$ and $|\vec{S}| = S$.)

We briefly summarize the essential properties of $\hat{S}$ and $\hat{L}$ needed for our discussion. In the (x, y, z) coordinate system, $\hat{L}$ has three components $\hat{L}_x$, $\hat{L}_y$ and $\hat{L}_z$. In most calculations we use $\hat{L}_z$ and the ladder operators $\hat{L}_+$ and $\hat{L}_-$ defined as $\hat{L}_\pm = \hat{L}_x \pm i\hat{L}_y$. When acted on the orbital state, they change $|L, L_z\rangle$ as

$$\begin{aligned}&\hat{L}_z|L, L_z\rangle \propto |L, L_z\rangle, \quad \hat{L}_\pm|L, L_z\rangle \propto |L, L_z \pm 1\rangle,\\ &\hat{L}_z|L, 0\rangle = 0, \quad \hat{L}_\pm|L, \pm L\rangle = 0\end{aligned} \tag{1a}$$

Namely, $\hat{L}_z$ does not change $L_z$, but $\hat{L}_+$ raises $L_z$ by 1 while $\hat{L}_-$ lowers $L_z$ by 1. If an independent coordinate (x′, y′, z′) is employed for the spin $\hat{S}$, the z′ direction is the preferred spin orientation by convention. The latter is specified with respect to the (x, y, z) coordinate by defining the polar angles θ and ϕ as depicted in **Fig. 2**. The three components of $\hat{S}$ in the (x′, y′, z′) coordinate are $\hat{S}_{x'}$, $\hat{S}_{y'}$ and $\hat{S}_{z'}$. Calculations involving $\hat{S}$ are carried out in terms of $\hat{S}_{z'}$ and the ladder operators $\hat{S}_{\pm'}$ defined by $\hat{S}_{\pm'} = \hat{S}_{x'} \pm i\hat{S}_{y'}$. In the notation for the spin state $|S, S_{z'}\rangle$ specified by two quantum numbers S and $S_{z'}$, the up-spin state of a single electron is described by $|\uparrow\rangle = |1/2, +1/2\rangle$, and the down-spin state by $|\downarrow\rangle = |1/2, -1/2\rangle$. When acted on the spin state, the operators $\hat{S}_{z'}$ and $\hat{S}_{\pm'}$ modify $|S, S_{z'}\rangle$ as

$$\begin{aligned}&\hat{S}_z|S, S_{z'}\rangle \propto |S, S_{z'}\rangle, \quad \hat{S}_\pm|S, S_{z'}\rangle \propto |S, S_{z'} \pm 1\rangle,\\ &\hat{S}_\pm|S, \pm S\rangle = 0\end{aligned} \tag{1b}$$

In evaluating whether or not the SOC-induced interactions between different electronic states vanish, one needs to recall that the orbital states $|L, L_z\rangle$ are orthonormal, and so are the spin states $|S, S_{z'}\rangle$. That is,

$$\langle L, L_z | L, L'_z\rangle = \begin{cases}1, & \text{if } L_z = L'_z\\ 0, & \text{otherwise}\end{cases} \quad , \quad \langle S, S_z | S, S'_z\rangle = \begin{cases}1, & \text{if } S_z = S'_z\\ 0, & \text{otherwise}\end{cases} \tag{2}$$

**Selection rules for preferred spin-orientation**

Using the (x, y, z) and (x′, y′, z′) coordinates for $\hat{L}$ and $\hat{S}$, respectively, the SOC Hamiltonian $\hat{H} = \lambda \hat{S} \cdot \hat{L}$ is written as $\hat{H} = \hat{H}^0_{SO} + \hat{H}'_{SO}$,[2,7-9] where

$$\hat{H}^0_{SO} = \lambda \hat{S}_{z'} \left( \hat{L}_z \cos\theta + \frac{1}{2}\hat{L}_+ e^{-i\phi} \sin\theta + \frac{1}{2}\hat{L}_- e^{+i\phi} \sin\theta \right) \quad (3a)$$

$$= \lambda \hat{S}_{z'} (\hat{L}_z \cos\theta + \hat{L}_x \sin\theta\cos\phi + \hat{L}_y \sin\theta\sin\phi) . \quad (3b)$$

$$\hat{H}'_{SO} = \frac{\lambda}{2} (\hat{S}_{+'} + \hat{S}_{-'}) \left( -\hat{L}_z \sin\theta + \hat{L}_x \cos\theta\cos\phi + \hat{L}_y \cos\theta\sin\phi \right) \quad (3c)$$

Whether the preferred spin orientation is parallel to the local z-direction (||z) (of the $ML_n$ under consideration) or perpendicular to it (⊥z) can be answered by using the above expression. The SOC-induced interaction between two *d*-states, $\psi_i$ and $\psi_j$, involves the interaction energy $\langle \psi_i | \hat{H} | \psi_j \rangle$. For our discussion, it is necessary to know whether this integral is zero or not. Since the angular part of a *d*- or *p*-orbital is expressed in terms of products $|L, L_z\rangle\, |S, S_{z'}\rangle$, the evaluation of $\langle \psi_i | \hat{H} | \psi_j \rangle$ involves the spin integrals $\langle S, S'_{z'} | \hat{S}_{z'} | S, S_{z'} \rangle$ as well as the orbital integrals $\langle L, L'_z | \hat{L}_z | L, L_z \rangle$ and $\langle L, L'_z | \hat{L}_\pm | L, L_z \rangle$.

The SOC Hamiltonian $\hat{H}^0_{SO}$ allows interactions only between identical spin states, because $\langle \uparrow | \hat{S}_{z'} | \uparrow \rangle$ and $\langle \downarrow | \hat{S}_{z'} | \downarrow \rangle$ are nonzero. For two states, $\psi_i$ and $\psi_j$, of identical spin, we consider the cases when $|\Delta L_z| = 0$ or 1. Then, we find

$$\langle \psi_i | \hat{H}^0_{SO} | \psi_j \rangle \propto \begin{cases} \cos\theta, & \text{if } |\Delta L_z| = 0 \\ \sin\theta, & \text{if } |\Delta L_z| = 1 \end{cases} \quad (4)$$

For the $|\Delta L_z| = 0$ case, $\langle \psi_i | \hat{H}^0_{SO} | \psi_j \rangle$ is maximum at θ = 0°, i.e., when the spin has the ||z orientation. For the $|\Delta L_z| = 1$ case, $\langle \psi_i | \hat{H}^0_{SO} | \psi_j \rangle$ becomes maximum at θ = 90°, i.e., when the spin

has the $\perp z$ orientation. Under SOC $\psi_i$ and $\psi_j$ do not interact when $|\Delta L_z| > 1$, because $\langle \psi_i | \hat{H}_{SO}^0 | \psi_j \rangle = 0$ in such a case.

The total energy of *ML<sub>n</sub>* is lowered under SOC by the interactions of the filled *d*-states with the empty ones. Since the strength of SOC is very weak, these interactions can be described in terms of perturbation theory in which the SOC Hamiltonian is taken as perturbation with the split *d*-states of $ML_n$ as unperturbed states. (These *d*-states are often referred to as the crystal-field split *d*-states, just to indicate that magnetic ions under consideration are embedded in crystalline solids.) Then, the most important interaction of the occupied *d*-states with the unoccupied *d*-states is the one between the HOMO and the LUMO (with energies $e_{HO}$ and $e_{LU}$, respectively), and the associated energy stabilization ΔE is given by Eq. 5.[1]

$$\Delta E = \begin{cases} -\left|\langle HO | \hat{H}_{SO}^0 | LU \rangle\right|, & \text{if } e_{HO} = e_{LU} \\ -\dfrac{\left|\langle HO | \hat{H}_{SO}^0 | LU \rangle\right|^2}{|e_{HO} - e_{LU}|}, & \text{if } e_{HO} < e_{LU} \end{cases} \quad (5)$$

Thus, we obtain the predictions for the preferred spin orientation as summarized in **Table 1**. In general, the effect of a degenerate interaction is stronger than that of a nondegenerate interaction. A system with degenerate HOMO and LUMO has Jahn-Teller (JT) instability, and the degeneracy would be lifted if the associated JT distortion were to take place.[10,11]

**Degenerate perturbation and uniaxial magnetism**

For a certain metal ion *M*, the electron configuration of $ML_n$ has unevenly-filled degenerate level. For example, the hexagonal perovskites $Ca_3CoMnO_6$ [12,13] consist of $CoMnO_6$ chains in which $CoO_6$ trigonal prisms containing high-spin $Co^{2+}$ (S = 3/2, $d^7$) ions alternate with $MnO_6$ octahedra containing high-spin $Mn^{4+}$ (S = 3/2, $d^3$) ions by sharing their triangular faces (**Fig. 3a**). A proper description of the electronic structure of a magnetic ion requires the use of the spin-polarized electronic structure method,[14-16] in which the energy and shape of *d*-states are allowed to depend on their electron occupancies. Nevertheless, the essential qualitative features of the spin-polarized electronic structure of $ML_n$ can be simulated using the electronic structures derived from an effective one-electron Hamiltonian by simply shifting rigidly the set of up-spin states lower in energy with respect to that of the down-spin states. Thus, the *d*-states of the high-spin $Co^{2+}$ (S = 3/2, $d^7$) ion in each $CoO_6$ trigonal prism (**Fig. 3b**) can be described by the electron configuration, $(z^2)^2 < (xy, x^2-y^2)^3 < (xz, yz)^2$, in the one-electron picture.[10,11,17] Thus, the spin-polarized d-states of the high-spin $Co^{2+}$ is written as,

$$(z^2\uparrow)^1 < (xy\uparrow, x^2-y^2\uparrow)^2 < (xz\uparrow, yz\uparrow)^2 < (z^2\downarrow)^1 < (xy\downarrow, x^2-y^2\downarrow)^1 < (xz\downarrow, yz\downarrow)^0.$$

Due to the half-filled degenerate set $(xy\downarrow, x^2-y^2\downarrow)$, the HOMO and LUMO are degenerate with $|\Delta L_z| = 0$, so the preferred spin orientation is ||z, i.e., along the three-fold rotational axis of the trigonal prism.

Such a magnetic ion exhibits uniaxial magnetism (i.e., Ising magnetism), that is, it has a nonzero magnetic moment $\vec{\mu}$ only in one direction in coordinate space. In classical mechanical terms, $\vec{\mu}$ is a change of the total energy E with respect to the applied magnetic field $\vec{H}$, i.e., $\vec{\mu} = -\partial E / \partial \vec{H}$.[2] In quantum mechanical terms, this energy change is equivalent to the split of a total angular momentum doublet-state by the Zeeman interaction,[2,3,17] $\hat{H}_Z = \mu_B (\hat{L} + 2\hat{S}) \cdot \vec{H}$, where $\mu_B$ is the Bohr magneton. For a system with uniaxial magnetism, a doublet state of the total angular momentum state is split only when a magnetic field is applied along the axis of the n-fold ($n \geq 3$) rotational symmetry. To examine the selection rule leading to uniaxial magnetism, we consider the total angular momentum number J. The degenerate set {xy, $x^2 - y^2$} of the $Co^{2+}$ (S = 3/2, $d^7$) ion, filled with three electrons, is equivalent to the {$L_z = +2$, $L_z = -2$} set. Thus, it has the unquenched orbital angular momentum $\vec{L}$ (with magnitude L = 2). The orbital momentum $\vec{L}$ couples with the spin momentum $\vec{S}$ by the SOC, $\lambda\vec{S}\cdot\vec{L}$, leading to the total angular momentum $\vec{J} = \vec{L} + \vec{S}$. The resulting total angular momentum states are specified by the two quantum numbers J and $J_z$, i.e., by $|J, J_z\rangle$. To identify the ground state of the spin-orbit coupled state, it is important to notice that $\lambda < 0$ for an ion with more than half-filled *d*-shell, but $\lambda > 0$ for the one with less than half-filled *d*-shell.[2] If $\lambda < 0$, the lowest-energy of the $\lambda\vec{S}\cdot\vec{L}$ term results when $\vec{S}$ and $\vec{L}$ are in the same direction. If $\lambda > 0$, however, it results when $\vec{S}$ and $\vec{L}$ have the opposite directions. Consequently, for a magnetic ion with L and S, the total angular quantum number J for the spin-orbit coupled ground state is given by

$$\text{Ground doublet}: \ J = \begin{cases} L + S & \text{if } \lambda < 0 \\ L - S & \text{if } \lambda > 0 \end{cases} \qquad (6)$$

For $\lambda < 0$, the energy of the J-state increases as J decreases. However, the opposite is the case for $\lambda > 0$. Since $\lambda < 0$ for the $Co^{2+}$ ($d^7$, S = 3/2) ion, J = L + S = 2 + 3/2 = 7/2 for the ground doublet state. Using the notations $|J, J_z\rangle$ for total angular momentum states, the two components of the doublet are described by $|J,+J\rangle$ and $|J,-J\rangle$. The degeneracy of $|J,+J\rangle$ and $|J,-J\rangle$ is always lifted by a magnetic field applied along the ||z direction (i.e., parallel to the axis of the n-fold rotational symmetry). However, for a magnetic field applied along the $\perp$z direction, the degeneracy is lifted only if J = 1/2.[2] Thus, for magnetic ions with unquenched orbital momentum, we find

$$\text{Magnetism} = \begin{cases} \text{uniaxial, if } J > 1/2 \\ \text{isotropic, otherwise} \end{cases} \qquad (7)$$

Let us now examine the uniaxial magnetism that arises from metal ions at octahedral sites by considering the $FeO_6$ octahedra with high-spin $Fe^{2+}$ ($d^6$, S = 2) ions present in the oxide $BaFe_2(PO_4)_2$, the honeycomb layers of which are made up of edge-sharing $FeO_6$ octahedra. This oxide exhibits a two-dimensional Ising magnetism.[18] For our analysis of its uniaxial magnetism, it is convenient to take the z-axis along one three-fold rotational axis of an $ML_6$ octahedron (**Fig. 4a**).[19] Then the orbital character of the *d*-states changes such that the $3z^2–r^2$ state becomes one of the $t_{2g}$ set, while the $\{xy, x^2–y^2\}$ set mixes with the $\{xz, yz\}$ set to give the new sets $\{1e_x, 1e_y\}$ and $\{2e_x, 2e_y\}$ (**Fig. 4b**). The high-spin $Fe^{2+}$ ion has the $(t_{2g})^4(e_g)^2$ configuration, the $(t_{2g})^4$ configuration of which can be described by $\Psi_{Fe,1}$ or $\Psi_{Fe,2}$ shown below

$$\Psi_{Fe,1} = (1a)^1(1e_x, 1e_y)^3$$
$$\Psi_{Fe,2} = (1a)^2(1e_x, 1e_y)^2$$

The occupancy of the down-spin d-states for $\Psi_{Fe,1}$ and $\Psi_{Fe,2}$ are presented in **Fig. 5a** and **5b**, respectively. An energy-lowering through SOC is allowed by $\Psi_{Fe,1}$ because it has an unevenly filled degenerate state $(1e_x, 1e_y)$, but not by $\Psi_{Fe,2}$ because the $(1e_x, 1e_y)$ set is evenly filled. The $(1e_x, 1e_y)^3$ configuration of $\Psi_{Fe,1}$ is also expressed as

$$(1e_x, 1e_y)^3 = \left(\sqrt{\tfrac{2}{3}}(xy, x^2 - y^2)^3 - \sqrt{\tfrac{1}{3}}(xz, yz)^3\right). \qquad (8)$$

The orbital-unquenched state $(xy, x^2 - y^2)^3 \equiv (xy\uparrow, x^2 - y^2\uparrow)^2(xy\downarrow, x^2 - y^2\downarrow)^1$ leads to L = 2, but the state $(xz, yz)^3 \equiv (xz\uparrow, yz\uparrow)^2(xz\downarrow, yz\downarrow)^1$ to L = 1. The SOC constant $\lambda < 0$ for the $\Psi_{Fe,1}$ configuration of $Fe^{2+}$ so that, with S = 2 for the $Fe^{2+}$ ion, the ground doublet has J = L + S = 4 from the component $(xy, x^2–y^2)^3$ (L = 2), and J = 3 from $(xz, yz)^3$ (L = 1). In terms of the notation $\{J_z,$ $-J_z\}$ representing a spin-orbit coupled doublet set, the doublet $\{4,-4\}$ is more stable than $\{3,-3\}$, so the $(1e_x, 1e_y)^3$ configuration of $Fe^{2+}$ is expressed as

$$\text{For high spin } Fe^{2+}: (1e_x, 1e_y)^3 \equiv \{4,-4\}^2\{3,-3\}^1$$

With J = 3 for the singly-filled doublet, uniaxial magnetism is predicted for the high-spin $Fe^{2+}$ ion at an octahedral site with ||z spin orientation. Note that the $\Psi_{Fe,2}$ configuration (**Fig. 5b**) leads to $|\Delta L_z| = 1$ and hence the preference for the ⊥z spin orientation. In support of this analysis, DFT calculations show the orbital moment of the $Fe^{2+}$ ion to be ~1 $\mu_B$ (i.e., L ≈ 1).[20]

Using the classical term $\lambda\vec{S}\cdot\vec{L}$, one can predict the ||z spin orientation for magnetic ions with unquenched orbital momentum, as discussed in the supporting information (SI).

**Nondegenerate perturbation and weak magnetic anisotropy**

We now examine the preferred spin orientations of magnetic ions with nondegenerate HOMO and LUMO, several examples of which are presented in **Fig. 6**. The layered compound $SrFeO_2$ consists of $FeO_2$ layers made up of corner-sharing $FeO_4$ square planes containing high-spin $Fe^{2+}$ ($d^6$, S = 2) ions.[21] Corner-sharing $FeO_4$ square planes are also found in $Sr_3Fe_2O_5$, in which they form two-leg ladder chains.[22] The *d*-states of a $FeO_4$ square plane are split as in **Fig. 6a**,[23,24] so that the down-spin d-states have only the $3z^2-r^2\downarrow$ level filled, with the empty {xz↓, yz↓} set lying immediately above. Thus, between these HOMO and LUMO, with $|\Delta L_z| = 1$ so the preferred spin direction is ⊥z, i.e., parallel to the $FeO_4$ plane.[23,24]

A regular $MnO_6$ octahedron containing a high-spin $Mn^{3+}$ ($d^4$, S = 2) ion has JT instability and hence adopts an axially-elongated $MnO_6$ octahedron (**Fig. 6b**). Such JT-distorted $MnO_6$ octahedra are found in $TbMnO_3$ [25] and $Ag_2MnO_2$.[26,27] The neutron diffraction studies show that the spins of the $Mn^{3+}$ ions are aligned along the elongated Mn-O bonds.[25,27] With four unpaired electrons to fill the split *d*-states, the LUMO is the $x^2–y^2$↑ and the HOMO is the $3z^2–r^2$↑. Between these two states, $|\Delta L_z| = 2$ so that they do not interact under SOC. The closest-lying filled *d*-state that can interact with the LUMO is the xy↑. Now, $|\Delta L_z| = 0$ between the $x^2–y^2$↑ and xy↑ states, the preferred spin orientation is ||z, i.e., parallel to the elongated Mn-O bonds.[27,28]

The $NiO_6$ trigonal prisms containing $Ni^{2+}$ ($d^8$, S = 1) ions are found in the $NiPtO_6$ chains of $Sr_3NiPtO_6$,[29] which is isostructural with $Ca_3CoMnO_6$. Each $NiPtO_6$ chain consists of face-sharing $NiO_6$ trigonal prisms and $PtO_6$ octahedra. The $Pt^{4+}$ ($d^6$, S = 0) ions are nonmagnetic. As depicted in **Fig. 6c** for the down-spin *d*-states of $Ni^{2+}$ ($d^8$, S = 1), $|\Delta L_z| = 1$ between the HOMO and LUMO. Consequently, the preferred spin orientation of the $Ni^{2+}$ ($d^8$, S = 1) ion is ⊥z, i.e., perpendicular to the $NiPtO_6$ chain. This in agreement with our DFT calculations.[30] For the discussion of the ||z spin orientation of the $Ni^{2+}$ ions in $Sr_3NiIrO_6$, see SI.

**Magnetic anisotropy of spin-half systems**

First, we consider the magnetic ions with S = 1/2 in which the HOMO and LUMO of the crystal-field *d*-states are not degenerate. An axially-elongated $IrO_6$ octahedra containing low-spin $Ir^{4+}$ ($d^5$, S = 1/2) ions are found in the layered compound $Sr_2IrO_4$, in which the corner-sharing of the $IrO_6$ octahedra using the equatorial oxygen atoms forms the $IrO_4$ layers with the elongated Ir-O bonds perpendicular to the layer.[31-33] The neutron diffraction studies of $Sr_2IrO_4$ show that the $Ir^{4+}$ spins are parallel to the $IrO_4$ layer.[32,33] With the z-axis chosen along the elongated Ir-O bond, the $t_{2g}$ level of the $IrO_6$ octahedron is split into {xz, yz} < xy. With five *d*-electrons to fill the three levels, the down-spin states xz↓ and yz↓ are filled while the xy↓ state is empty, as depicted in **Fig.**

**7a**. Consequently, $|\Delta L_z| = 1$ between the HOMO and LUMO, so that the preferred spin orientation is $\perp z$. This is in agreement with experiment and our DFT calculations.[30]

$CuCl_2 \cdot 2H_2O$ is a molecular crystal made up of $CuCl_2(OH_2)_2$ complexes containing $Cu^{2+}$ ($d^9$, S = 1/2) ions, in which the linear O-Cu-O unit is perpendicular to the linear Cl-Cu-Cl unit (**Fig. 7b**).[34] The spins of the $Cu^{2+}$ ions are aligned along the Cu-O direction,[35] namely, the $Cu^{2+}$ ions have easy-plane anisotropy. The split down-spin *d*-states of $CuCl_2 \cdot 2H_2O$ show that the LUMO, $x^2{-}y^2\!\downarrow$ has the smallest energy gap with the HOMO, $xz\!\downarrow$ **(Fig. 7b**).[8] Since $|\Delta L_z| = 1$, the preferred spin orientation is $\perp z$. To see if the spin prefers the x- or y-direction in the xy-plane, we use Eq. 3b. The matrix elements $\langle \psi_i | \hat{L}_\mu | \psi_j \rangle$ of the angular momentum operators $\hat{L}_\mu (\mu = x, y, z)$ are nonzero only for the following { $\psi_i$ , $\psi_j$ } sets:[2]

$$\begin{array}{llllll} \text{For } \hat{L}_z: & \{xz, yz\}, & \{xy, x^2{-}y^2\}, & & \{x, y\} & \\ \text{For } \hat{L}_x: & \{yz, 3z^2{-}r^2\}, & \{yz, x^2{-}y^2\}, & \{xz, xy\}, & \{y, z\} & (9) \\ \text{For } \hat{L}_y: & \{xz, 3z^2{-}r^2\}, & \{xz, x^2{-}y^2\}, & \{yz, xy\}, & \{z, x\} & \end{array}$$

The only nonzero interaction between the LUMO$x^2{-}y^2\!\downarrow$ and the HOMO $xz\!\downarrow$ under SOC is the term $\langle x^2 - y^2 | \hat{L}_y | xz \rangle$ involving $\hat{L}_y$ . Eq. 3b shows that this term comes with angular dependency of $\sin\theta\sin\phi$, which is maximized when $\theta = 90°$ and $\phi = 90°$. Thus, the preferred spin orientation of $CuCl_2(OH_2)_2$ is along the y-direction, namely, along the Cu-O bonds.[8]

In $CuCl_2$,[36,37] $CuBr_2$ [38] and $LiCuVO_4$,[39] the square planar $CuL_4$ units (*L* = Cl, Br, O) share their opposite edges to form $CuL_2$ ribbon chains (**Fig. 8a**). The split *d*-states in the $CuL_2$ ribbon chains of $CuCl_2$, $CuBr_2$ and $LiCuVO_4$ can be deduced by examining their projected density of states (PDOS) plots. Analyses of these plots can be best described by the effective sequence of the down-spin *d*-states shown in Eq. 10a.[8]

$$(3z^2{-}r^2\!\downarrow)^1(xy\!\downarrow)^1(xz\!\downarrow, yz\!\downarrow)^2(x^2{-}y^2\!\downarrow)^0 \text{ for a } CuL_4 \text{ of a } CuL_2 \text{ ribbon chain} \quad (10a)$$

$(3z^2{-}r^2{\downarrow})^1(xz{\downarrow}, yz{\downarrow})^2(xy{\downarrow})^1(x^2{-}y^2{\downarrow})^0$ for an isolated Cu$L_4$ square plane (10b)

Consequently, the interaction of the LUMO $x^2$-$y^2\downarrow$ with the HOMO (xz↓, yz↓) will lead to the ⊥z spin orientation for the $Cu^{2+}$ ions of the Cu$L_2$ ribbon chains.[8] This down-spin *d*-state sequence is different from the corresponding one expected for an isolated Cu$L_4$ square plane (shown in Eq. 10b). This is due to the orbital interactions between adjacent Cu$L_4$ square planes in the Cu$L_2$ ribbon chain, in particular, the direct metal-metal interactions involving the xy orbitals through the shared edges between adjacent Cu$L_4$ square planes.

Now we consider the magnetic ions with S = 1/2 whose HOMO and LUMO are degenerate. $Sr_3NiIrO_6$ [40] is isostructural with $Ca_3CoMnO_6$, and its $NiIrO_6$ chains are made up of face-sharing $IrO_6$ octahedra and $NiO_6$ trigonal prisms. Each $NiO_6$ trigonal prism has a $Ni^{2+}$ ($d^8$, S = 1) ion, and each $IrO_6$ octahedron a low-spin $Ir^{4+}$ ($d^5$, S = 1/2) ion. Magnetic susceptibility and magnetization measurements[30,41] indicate that $Sr_3NiIrO_6$ has uniaxial magnetism with the spins of both $Ni^{2+}$ and $Ir^{4+}$ ions aligned along the chain direction. Neutron diffraction measurements show that in each chain the spins of adjacent $Ni^{2+}$ and $Ir^{4+}$ ions are antiferromagnetically coupled.[41] The low-spin $Ir^{4+}$($d^5$, S =1/2) ion has the configuration $(t_{2g})^5$, which can be represented by $\Psi_{Ir,1}$ or $\Psi_{Ir,2}$

$$\Psi_{Ir,1} = (1a)^2(1e_x, 1e_y)^3$$
$$\Psi_{Ir,2} = (1a)^1(1e_x, 1e_y)^4$$

The occupancies of the down-spin *d*-states for $\Psi_{Ir,1}$ and $\Psi_{Ir,2}$ are given as depicted in **Fig. 5c** and **5d**, respectively. It is $\Psi_{Ir,1}$, not $\Psi_{Ir,2}$, that can lower energy under SOC. The $(1e_x, 1e_y)^3$ part of $\Psi_{Ir,1}$ can be rewritten as in Eq. 8. For the low-spin $Ir^{4+}$, $\lambda < 0$, because the $t_{2g}$-shell is more than half-filled.[30] With S = 1/2, we have J = L + S = 5/2 from $(xy, x^2{-}y^2)^3$, and 3/2 from $(xz, yz)^3$. Thus, the $(1e_x, 1e_y)^3$ configuration of $Ir^{4+}$ is expressed as

For low spin $Ir^{4+}$ : $(1e_x, 1e_y)^3 \equiv \{5/2, -5/2\}^2 \{3/2, -3/2\}^1$

The singly-filled doublet has J = 3/2, so uniaxial magnetism is predicted with the spin orientation along the ||z direction. This explains why the S=1/2 ion $Ir^{4+}$ ion exhibits a strong magnetic anisotropy with the preferred spin direction along the z-axis.

The $Os^{7+}$ ($d^1$, S = 1/2) ion of each $OsO_6$ octahedron in the double-perovskite $Ba_2NaOsO_6$ [42-44] have degenerate HOMO and LUMO. The $(t_{2g})^1$ configuration that can have energy-lowering through SOC is given by $\Psi_{Os,1} = (1e_x, 1e_y)^1$ (**Fig. 9a**), which is equivalent to

$$\begin{aligned}\Psi_{Os,1} &= \sqrt{\tfrac{2}{3}}(xy, x^2 - y^2)^1 - \sqrt{\tfrac{1}{3}}(xz, yz)^1 \\ &\equiv \{1/2, -1/2\}^1\end{aligned}$$

For the $Os^{7+}$ ions, the SOC constant is positive ($\lambda > 0$), because its *d*-shell is less than half-filled. Thus, the spin-orbit coupled ground doublet has J = L – S. For the ground doublet resulting from $\Psi_{Os,1}$ shows J = 3/2 from $(xy, x^2 - y^2)^1$ (L = 2), but J = 1/2 from $(xz, yz)^1$ (L = 1). Of these two, the ground doublet has J = 1/2 since $\lambda > 0$. The isotropic magnetic properties of $Ba_2NaOsO_6$ are explained since J = 1/2 for the ground doublet.[17] The preferred spin orientation predicted by the value of $|\Delta L_z| = 0$ is the ||z direction. Apparently, this seems inconsistent with the prediction based on J = 1/2. However, a given $ML_6$ octahedron has four different $C_3$ rotational axes, all of which provide equally valid local z-directions. As long as the shape of the $ML_6$ octahedron remains ideal, all directions are equally valid, namely, the system is isotropic.

Let us consider the spin orientation of the S=1/2 ions $V^{4+}$ ($d^1$) in the $VO_6$ octahedra of $R_2V_2O_7$ (R = rare earth),[45] in which each $VO_6$ octahedron is axially compressed along the direction of its local three-fold rotational axis so that its $t_{2g}$ state is split into the 1a < 1e pattern (**Fig. 9b**). With the local z-axis along the three-fold rotational axis of $VO_6$, the HOMO is the 1a↑ state, which is represented by $3z^2–r^2\uparrow$, which interacts with the LUMO $1e\uparrow = (1e_x\uparrow, 1e_y\uparrow)$ states under SOC through their $(xz\uparrow, yz\uparrow)$ components. Consequently, $|\Delta L_z| = 1$ and the preferred spin orientation would be $\perp z$. However, the observed spin orientation is $\|z$,[46] which has also been confirmed by DFT calculations.[47] This finding is explained if the $V^{4+}$ ion has some uniaxial magnetic character despite that the HOMO and LUMO are not degenerate. For the latter to be true, the true ground state of each $V^{4+}$ ion in $R_2V_2O_7$ should be a "contaminated state" 1a′, which has some contributions of the 1e and 2e character of its isolated $VO_6$ octahedron, namely,

$$|1a'\rangle \propto |1a\rangle + \gamma|1e\rangle + \delta|2e\rangle \quad (11)$$

where γ and δ are small mixing coefficients. This is possible because each $VO_6$ octahedron present in $R_2V_2O_7$ has a lower symmetry than does an isolated $VO_6$ octahedron. The $VO_6$ octahedra are corner-shared to form a tetrahedral cluster (**Fig. 9b**), and such tetrahedral clusters further share their corners to form a pyroclore lattice (**Fig. 9c**). Indeed, the PDOS plots for the up-spin *d*-states of the $V^{4+}$ ions in $R_2V_2O_7$ show the presence of slight contributions of the 1e and 2e states to the occupied 1a state (**Fig. 9d**).

As reviewed above, the magnetic anisotropy S = 1/2 ions can be strong, weak or vanishing depending on how their split *d*-states interact among themselves under SOC. For a long time there has been a blind faith that S = 1/2 ions in solids or molecules must have no magnetic anisotropy arising from SOC. A classic example is found in the six-decade-old study of $CuCl_2\cdot 2H_2O$,[48] which explored the origin of the observed $Cu^{2+}$-spin orientation after first dismissing SOC as a possible cause. Unfortunately, this "spin-half syndrome" still remains unabated. J = 1/2 ions have no magnetic anisotropy, but this is not necessarily true for S = 1/2 ions.

## Ligand-controlled spin orientation

For the $CuBr_4$ square planes of $CuBr_2$ ribbon chain,[38] the $CuBr_5$ square pyramids of $(C_5H_{12}N)CuBr_3$,[49,50] and the $CrI_6$ octahedra of the layered compound $CrI_3$,[51] the ligand *L* is heavier than *M*, so the SOC between two *d*-states of $ML_n$ results more from the SOC-induced interactions between the *p*-orbitals of the ligands *L* rather than from those between the *d*-orbitals of *M*. We clarify this point by considering a square planar $ML_4$ using the coordinate system of **Fig. 9a**. The metal and ligand contributions in the yz, xy and $x^2{-}y^2$ states of $ML_4$ are shown in **Fig. 9b-d**, respectively. The SOC-induced interaction between different *d*-states can occur by the SOC of *M*, and also by that of each ligand *L*. The interaction between the z and {x, y} orbitals at each *L* has $|\Delta L_z| = 1$, leading to the ⊥z spin orientation. In contrast, the interaction between the x and y orbitals at each *L* has $|\Delta L_z| = 0$, leading to the ||z spin orientation (**Table 1**). When the ligand *L* is much heavier than the metal *M*, the SOC constant λ of *L* is greater than that of *M*. Furthermore, such ligands *L* possess diffuse and high-lying *p*-orbitals, which makes the magnetic orbitals of $ML_n$ dominated by the ligand *p*-orbitals and also makes the *d*-states of $ML_n$ weakly split. This makes the SOC effect in $ML_n$ dominated by the ligands (see SI for further discussion).

**High-spin d$^5$ systems**

High-spin d$^5$ transition-metal ions with S = 5/2 possess a small nonzero orbital momentum $\delta\vec{L}$ and exhibit weakly preferred spin orientations. For such a magnetic ion, the SOC-induced HOMO-LUMO interaction should be based on the $\hat{H}'_{SO}$ term (Eq. 3c), because the HOMO and LUMO occur from different spin states. The comparison of Eq. 3b with Eq. 3c reveals that the predictions concerning the ||z vs. ⊥z spin orientation from the term $\hat{H}'_{SO}$ are exactly opposite to those from the term $\hat{H}^0_{SO}$.

## Conclusion

The uniaxial magnetism and preferred spin orientations of magnetic ions can be reliably predicted by analyzing the $|\Delta L_z|$ values associated with their HOMO-LUMO interactions induced by SOC.

## ASSOCIATED CONTENT

### Supporting Information

The Supporting Information is available free of charge on the ACS Publications website:

Vector analysis of uniaxial magnetism, effect of spin exchange on spin orientation, and weak anisotropy of high-spin $Fe^{2+}$ ions in $FeCl_6$ octahedra.

## AUTHOR INFORMATION

### Corresponding Author

*E-mail: mike_whangbo@ncsu.edu

### Notes

The authors declare no competing financial interest.

### Biographies

**Myung-Hwan Whangbo** studied at Seoul National University receiving his bachelor's and master's degrees in 1968 and 1970, respectively, and at Queen's University receiving his Ph. D. in 1974. After postdoctoral studies at Queen's and Cornell Universities, he started his academic career at North Carolina State University in 1978, where he is a Distinguished Professor.

**Elijah E. Gordon** received his B. S. degree from North Carolina State University in 2013, and has been studying toward his Ph. D. degree under the supervision of M.-H. Whangbo.

**Hongjun Xiang** received his Ph. D. degree from University of Science and Technology of China in 2006. After postdoctoral studies at North Carolina State University and at National Renewable Energy Laboratory, he has been a Professor of Physics at Fudan University since 2009.

**Hyun-Joo Koo** studied at Sungkyunkwan University receiving her Ph. D. in 1997. After a postdoctoral study at North Carolina State University, she has been a Professor of Chemistry at Kyung Hee University, Seoul since 2004.

**Changhoon Lee** received his Ph. D. degreebfrom Wonkwang University in 2005. After a postdoctoral study at North Carolina State University, he has been working at POSTECH as a Research Assistant Professor of Chemistry since 2011.

## ACKNOWLEDGMENTS

This work was supported by the computing resources of the NERSC Center and the HPC Center of NCSU. H.J.X. thanks NSFC, FANEDD, NCET-10-0351, Research Program of Shanghai Municipality and MOE, and the Special Funds for Major State Basic Research. C.L. thanks National Research Foundation of Korea for the Basic Science Research Program, NRF-2013R1A1A2060341.

Table 1. The preferred spin orientations of magnetic ions predicted using the $|\Delta L_z|$ values associated with the SOC-induced HOMO-LUMO interactions

| Spin orientation | Requirement | Interactions between |
|---|---|---|
| $\parallel z$ | $\lvert\Delta L_z\rvert = 0$ | xz and yz<br>xy and $x^2-y^2$<br>x and y |
| $\perp z$ | $\lvert\Delta L_z\rvert = 1$ | $\{3z^2-r^2\}$ and {xz, yz}<br>{xz, yz} and $\{xy, x^2-y^2\}$<br>z and {x, y} |

**Figure captions**

Figure 1. The minimum difference in the magnetic quantum numbers, $|\Delta L_z|$, (a) between pairs of *d*-orbitals and (b) between pairs of *p*-orbitals.

Figure 2. The polar angles θ and ϕ defining the preferred orientation of the spin (i.e., the z′-axis) with respect to the (x, y, z) coordinate used to describe the orbital.

Figure 3. (a) A schematic view of an isolated $CoMnO_6$ chain of $Ca_3CoMnO_6$, which is made up of the $CoO_6$ trigonal prisms containing high-spin $Co^{2+}$ ($d^7$, S = 3/2) ions and the $MnO_6$ octahedra containing high-spin $Mn^{4+}$ ($d^3$, S = 3/2) ions. (b) The occupancy of the down-spin *d*-states for a high-spin $Co^{2+}$ ion in an isolated $CoO_6$ trigonal prism.

Figure 4. (a) The coordinate of an $ML_6$ octahedron with the z-axis taken along one of the four three-fold rotational axes. (b) The components of the *d*-orbitals of an $ML_6$ octahedron in the coordinate system of (a).

Figure 5. The down-spin electron configurations of a high-spin $Fe^{2+}$ ($d^6$, S = 2) at an octahedral site that induce (a) uniaxial magnetism and (b) no uniaxial magnetism, and those of a low-spin $Ir^{4+}$ ($d^5$, S = 1/2) ion at an octahedral site that induce (c) uniaxial magnetism and (d) no uniaxial magnetism.

Figure 6. (a) The down-spin electron configuration of a high-spin $Fe^{2+}$ ($d^6$, S = 2) at a square-planar site. (b) The up-spin electron configuration of a high-spin $Mn^{3+}$ ($d^4$, S = 2) at an axially-

elongated octahedral site. (c) The down-spin electron configuration of a $Ni^{2+}$ ($d^8$, S = 1) ion at a trigonal prism site.

Figure 7. (a) The structure and the down-spin *d*-states of a $CuCl_2(OH_2)_2$ complex: blue circle = Cu, green circle = Cl, red circle = O, and white circle = H. (b) The down-spin electron configuration of a low-spin $Ir^{4+}$ ($d^5$, S = 1/2) ion at an axially-elongated octahedral site.

Figure 8. (a) The $CuL_2$ ribbon chain made up of edge-sharing $CuL_4$ square planes. The contributions of the metal *d*- and the ligand *p*-orbitals in the (b) yz, (c) xy and (d) $x^2{-}y^2$ states of a $CuL_4$ square plane.

Figure 9. (a) The electron configurations of an $Os^{7+}$ ($d^1$, S = 1/2) ion at an octahedral site. (b) The split $t_{2g}$ state of a $V^{4+}$ ($d^1$, S = 1/2) ion at each $VO_6$ octahedron in $R_2V_2O_7$ (R = rare earth). (c) The pyrochlore lattice of the $V^{4+}$ ions in $R_2V_2O_7$. (d) The PDOS plots for the up-spin *d*-states of the $V^{4+}$ ions in $R_2V_2O_7$, which shows that slight contributions of the 1e and 2e states exist in the region of the occupied 1a state.

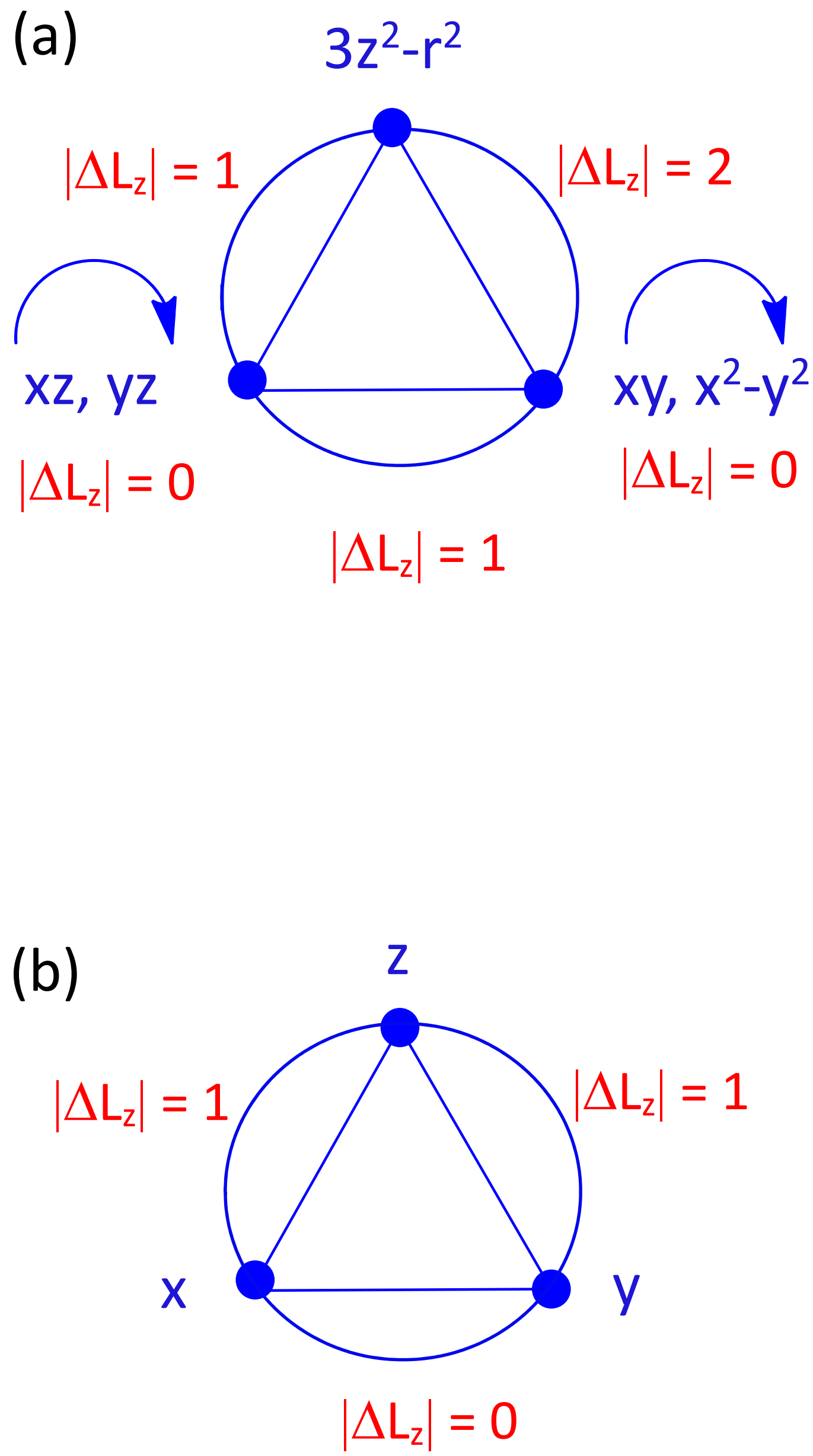
(a)
3z2-r2
|ΔLz| = 1
|ΔLz| = 2
xz, yz
xy, x2-y2
|ΔLz| = 0
|ΔLz| = 0
|ΔLz| = 1
(b)
z
|ΔLz| = 1
|ΔLz| = 1
x
y
|ΔLz| = 0

Figure 1.

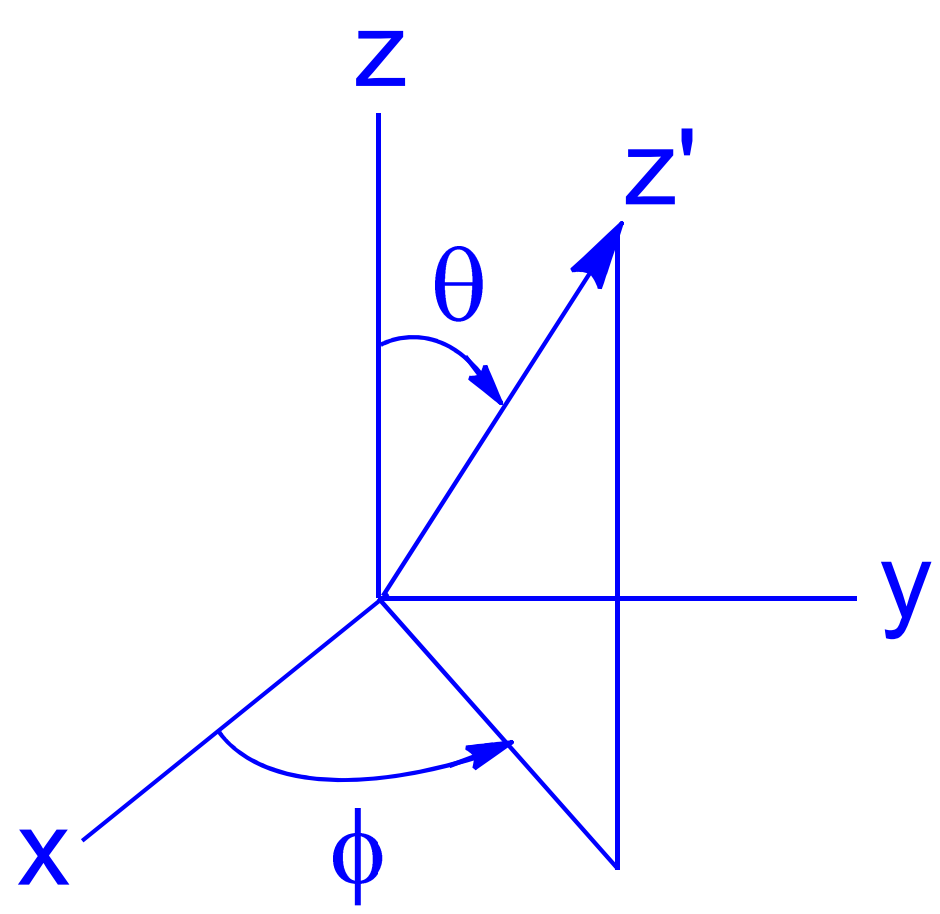

Figure 2

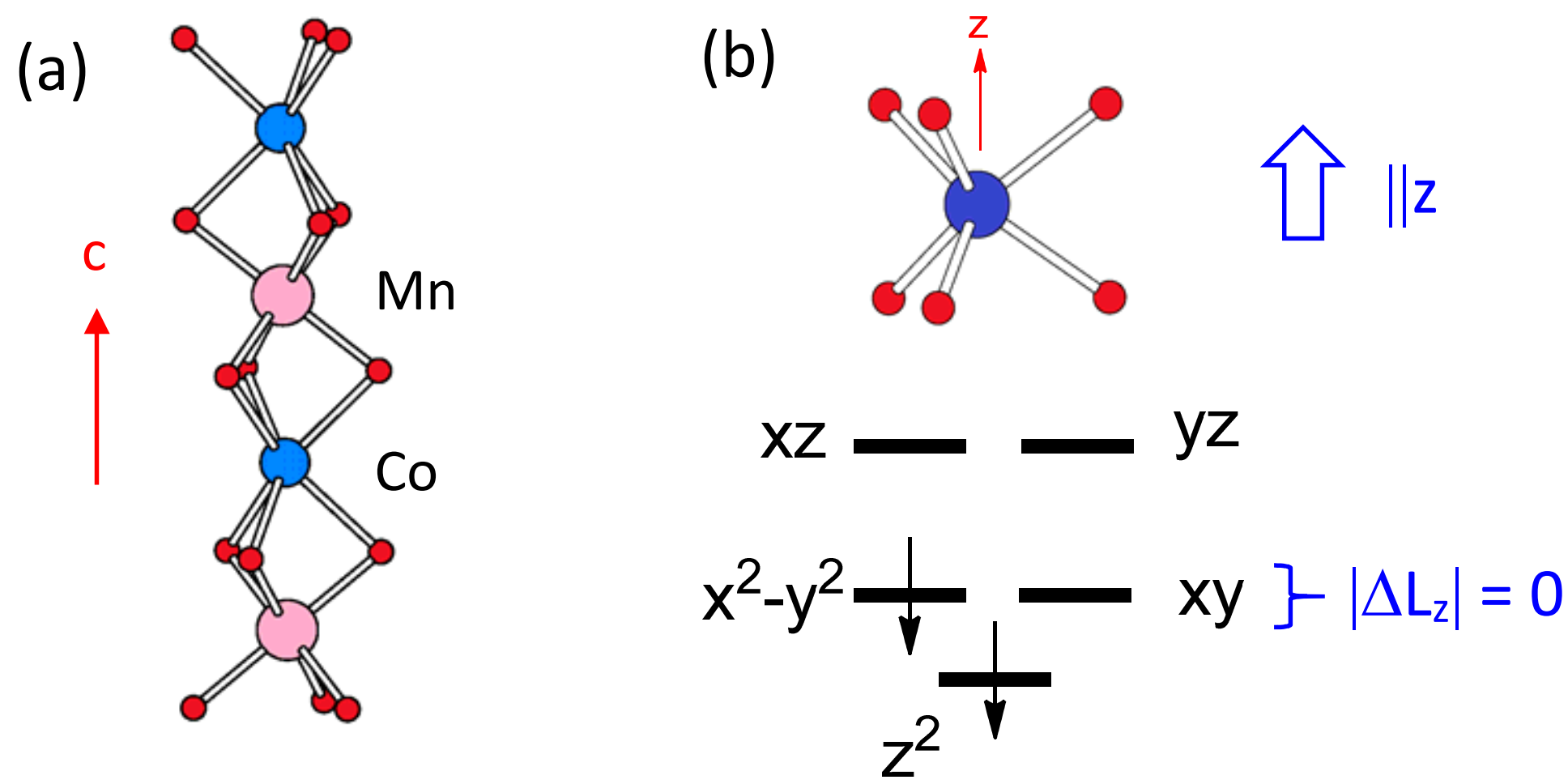

Figure 3

(a)

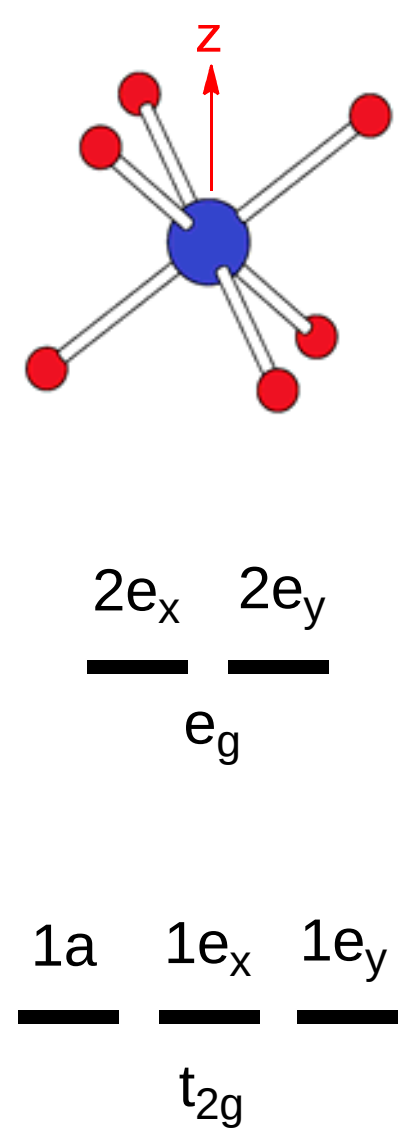

(b)

$$2e_x \equiv \sqrt{\frac{1}{3}}xy + \sqrt{\frac{2}{3}}xz$$

$$2e_y \equiv \sqrt{\frac{1}{3}}(x^2 - y^2) + \sqrt{\frac{2}{3}}yz$$

$$1a \equiv 3z^2 - r^2$$

$$1e_x \equiv \sqrt{\frac{2}{3}}xy - \sqrt{\frac{1}{3}}xz$$

$$1e_y \equiv \sqrt{\frac{2}{3}}(x^2 - y^2) - \sqrt{\frac{1}{3}}yz$$

Figure 4

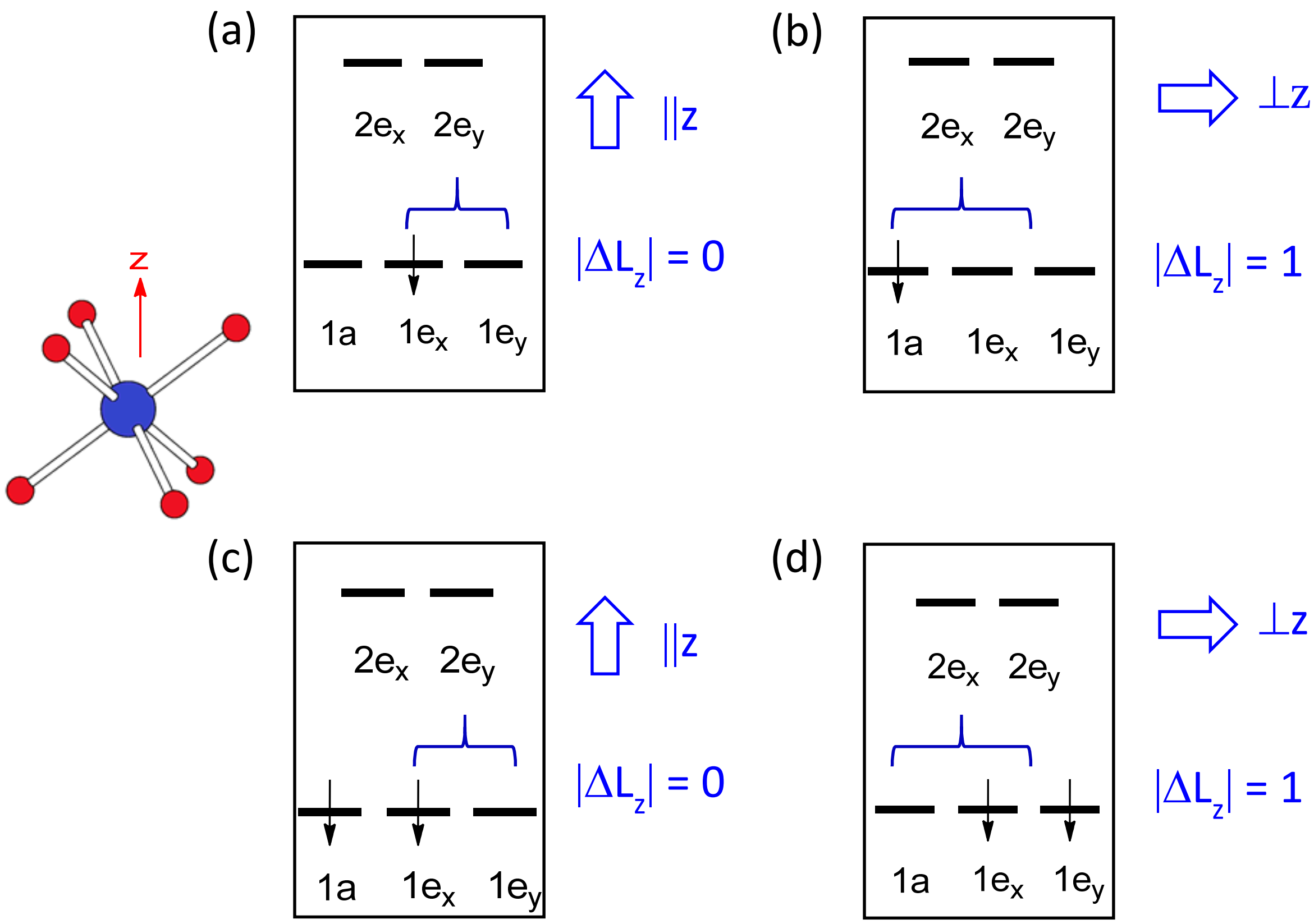
(a)
2ex 2ey
1a 1ex 1ey
||z
|ΔLz| = 0
(b)
2ex 2ey
1a 1ex 1ey
⊥z
|ΔLz| = 1
z
(c)
2ex 2ey
1a 1ex 1ey
||z
|ΔLz| = 0
(d)
2ex 2ey
1a 1ex 1ey
⊥z
|ΔLz| = 1

Figure 5

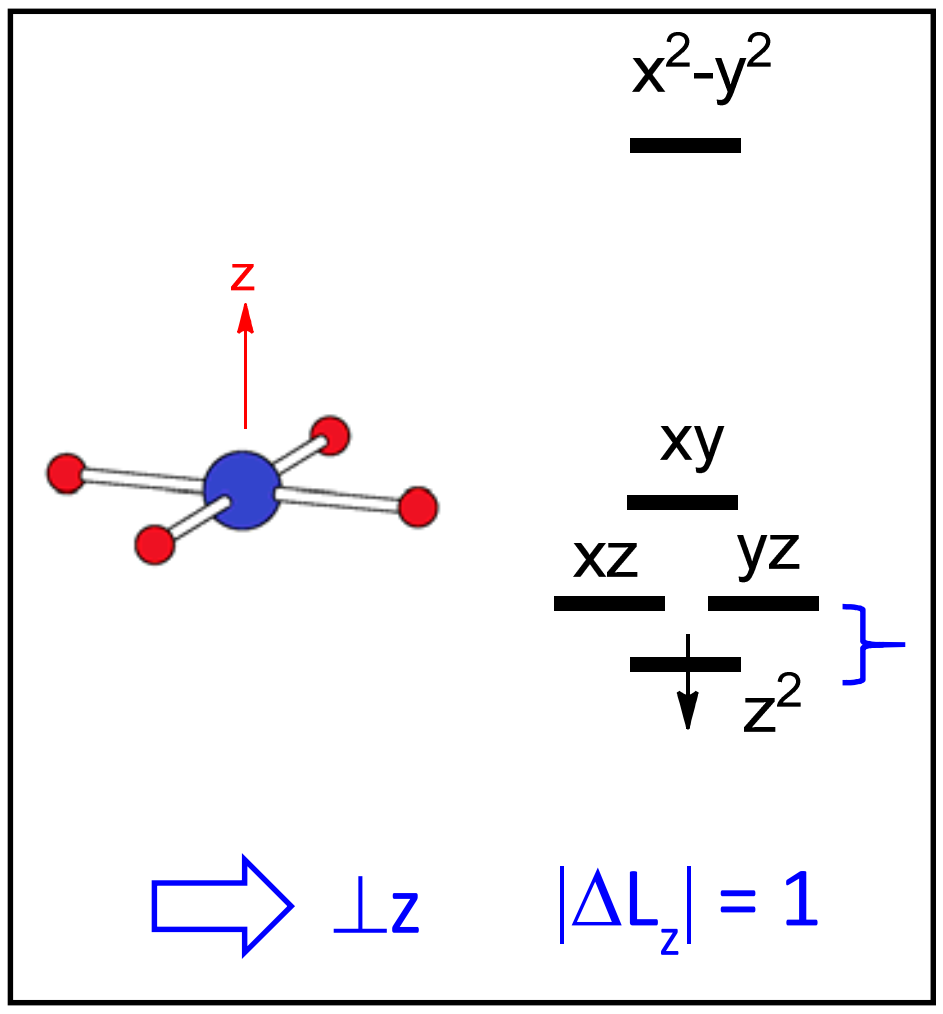

(a) $Fe^{2+}$ ($d^6$, S = 2)

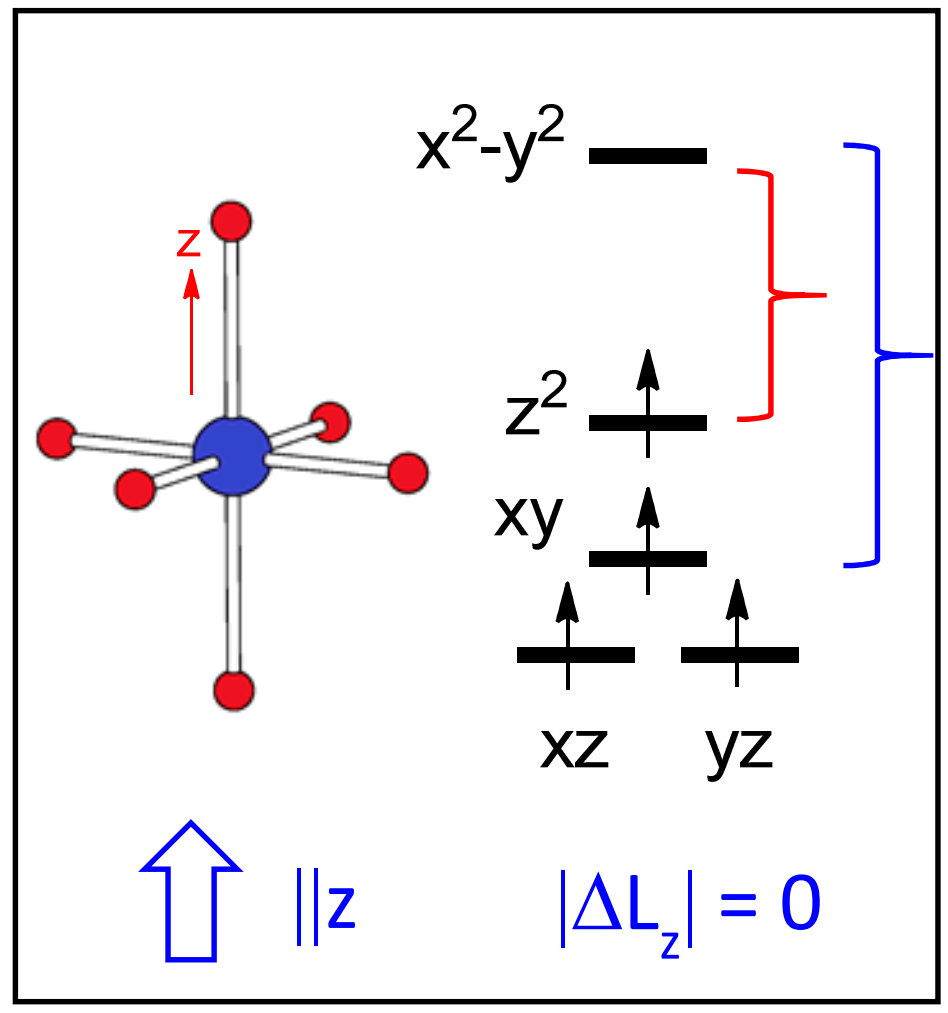

(b) $Mn^{3+}$ ($d^4$, S = 2)

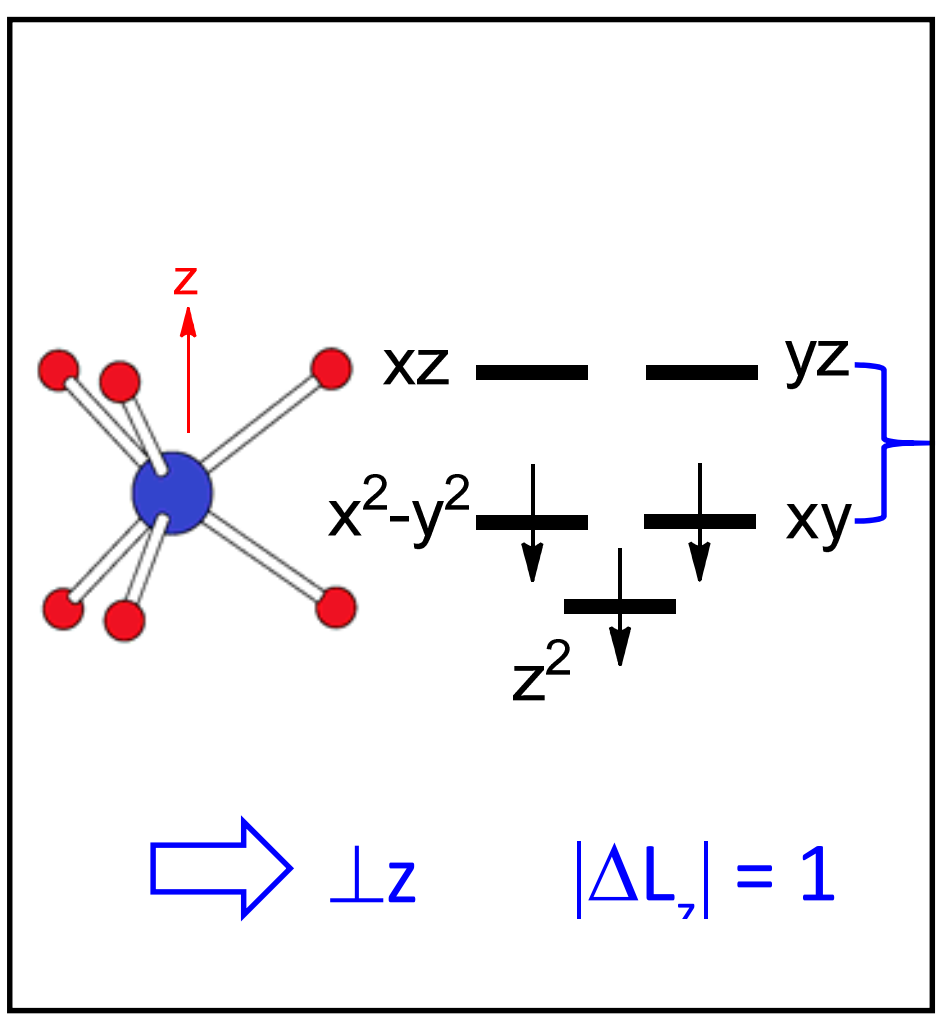

(c) $Ni^{2+}$ ($d^8$, S = 1)

Figure 6

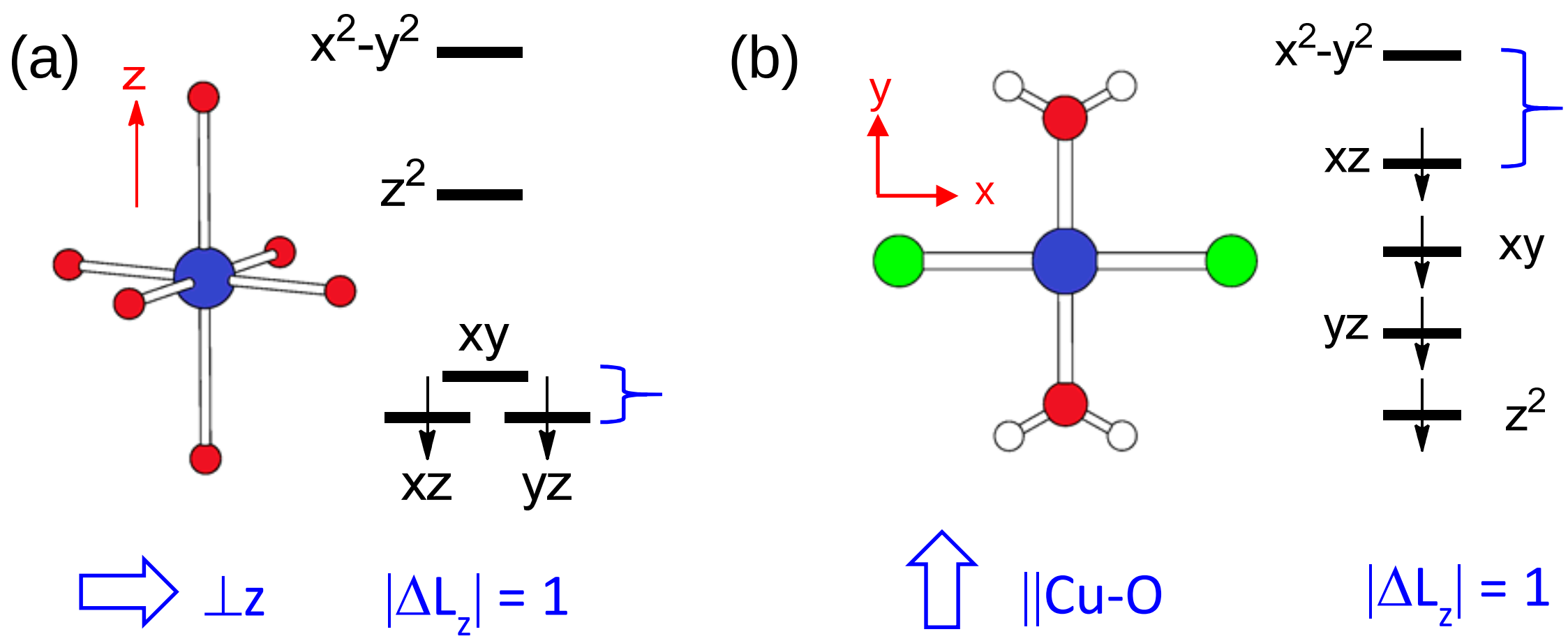

Figure 7

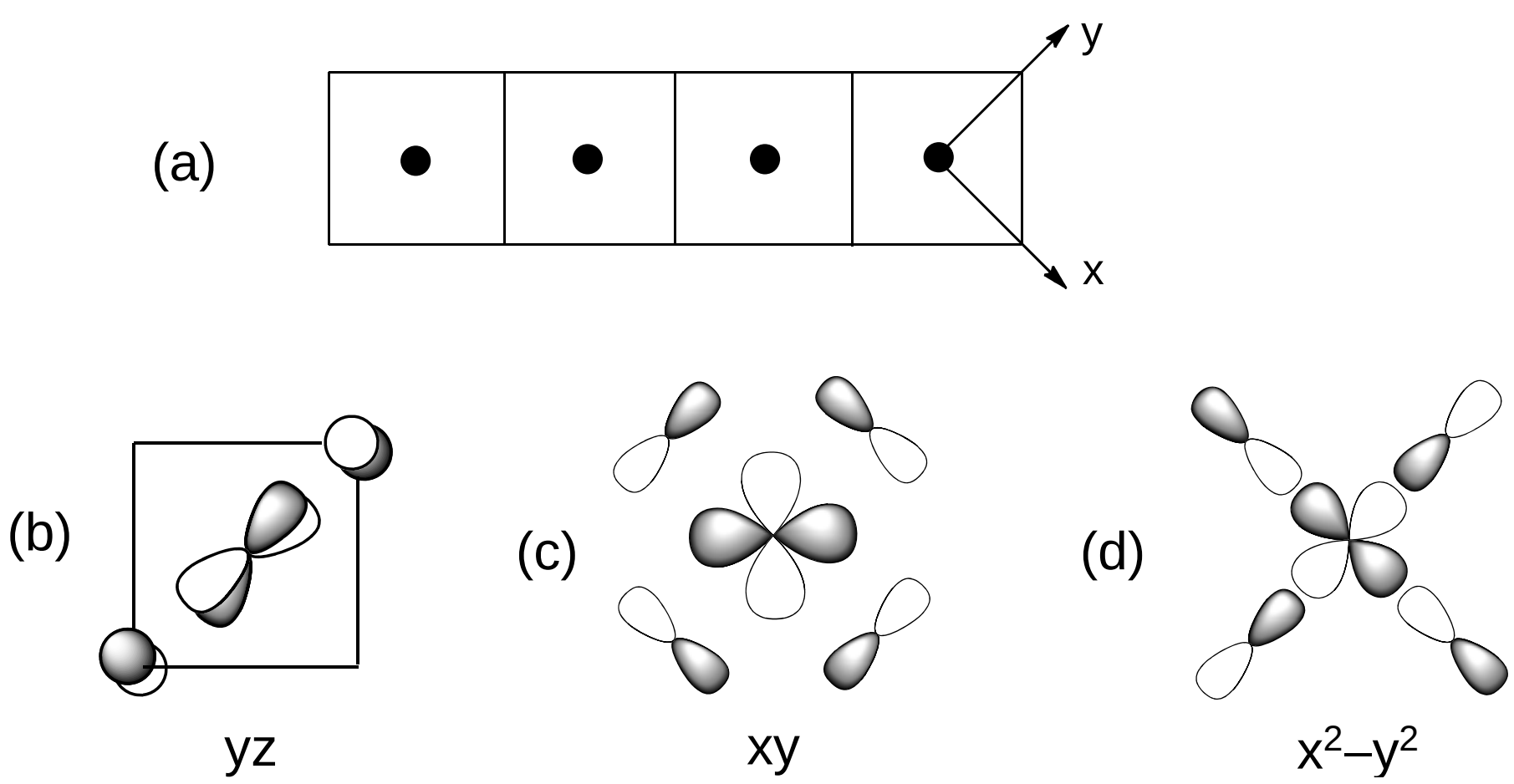

Figure 8

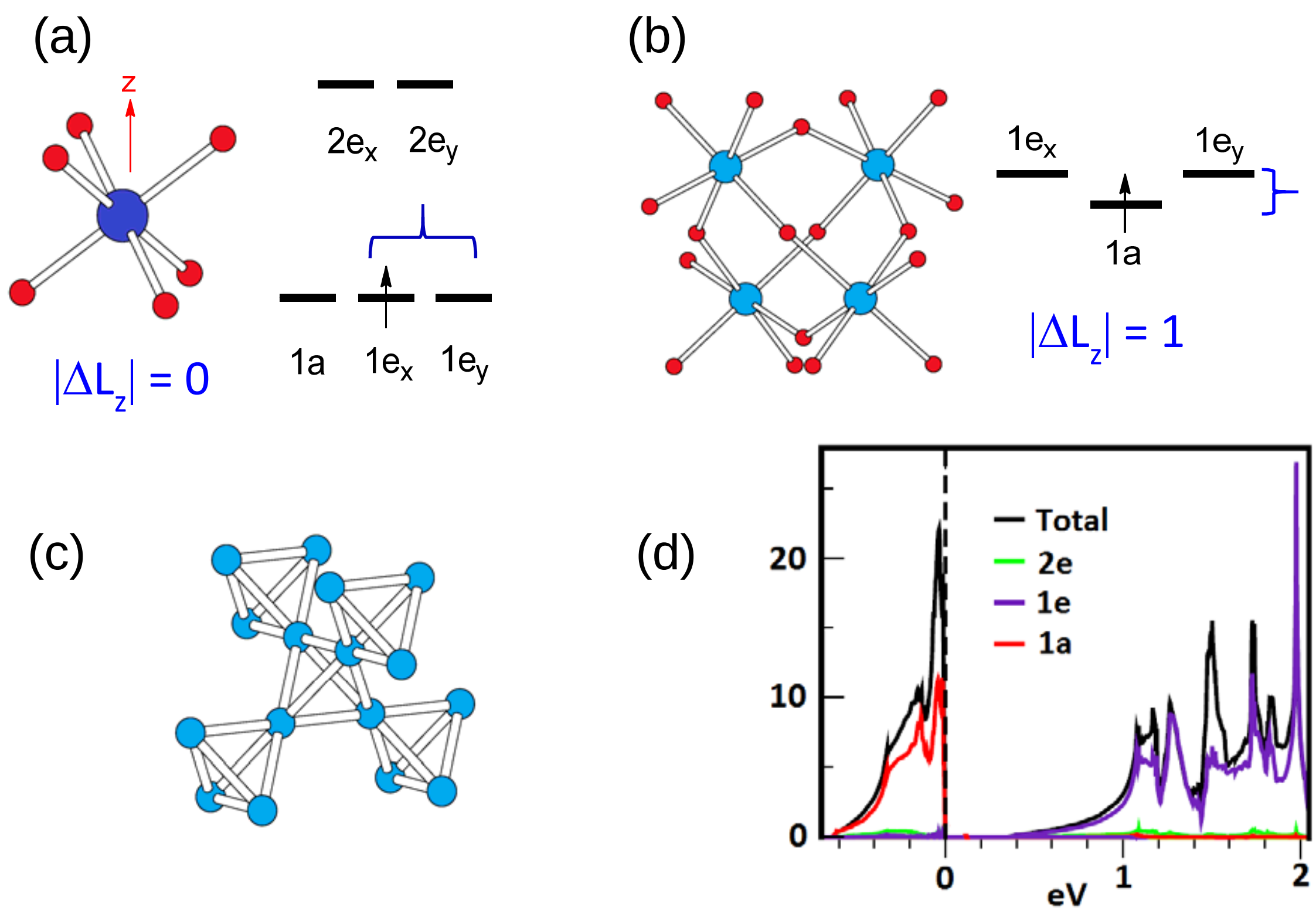
(a)
z
2ex 2ey
1a 1ex 1ey
|ΔLz| = 0
(b)
1ex 1ey
1a
|ΔLz| = 1
(c)
(d)
Total
2e
1e
1a
0
10
20
0
1
2
eV

Figure 9

## Graphics for the Conspectus

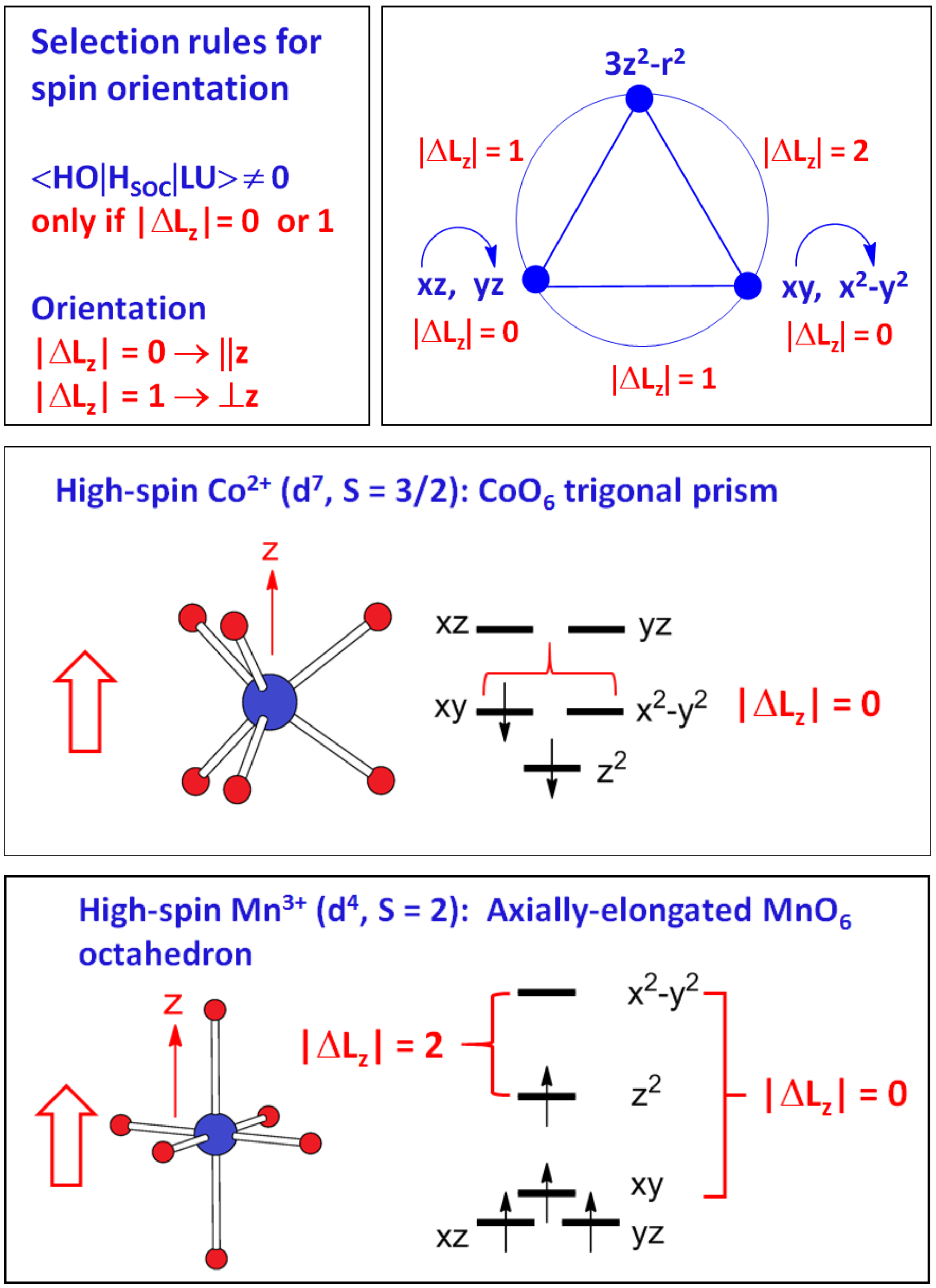

Supporting Information

for

**Prediction of spin orientations in terms of HOMO-LUMO interactions using spin-orbit coupling as perturbation**

Myung-Hwan Whangbo[a,*], Elijah E. Gordon[a], Hongjun Xiang[b], Hyun-Joo Koo[c], and Changhoon Lee[d]

[a] Department of Chemistry, North Carolina State University, Raleigh, NC 27695-8204, USA

[b] Key Laboratory of Computational Physical Sciences (Ministry of Education), State Key Laboratory of Surface Physics, Collaborative Innovation Center of Advanced Microstructures, and Department of Physics, Fudan University, Shanghai 200433, P. R. China

[c] Department of Chemistry and Research Institute for Basic Science, Kyung Hee University, Seoul 130-701, Republic of Korea

[d] Department of Chemistry, Pohang University of Science and Technology, Pohang 790-784, Korea

**1. Classical vector analysis of uniaxial magnetism**

It is of interest to consider the above discussion from the viewpoint of the classical SOC term, $\lambda\vec{S}\cdot\vec{L}$. A magnetic ion with unevenly-filled degenerate *d*-set has the unquenched orbital momentum $\vec{L}$, which is aligned along the rotational axis. Therefore, if $\lambda < 0$, the maximum energy gain from $\lambda\vec{S}\cdot\vec{L}$ occurs when the spin momentum $\vec{S}$ is in the same direction of $\vec{L}$. If $\lambda > 0$, however, the minimum energy of $\lambda\vec{S}\cdot\vec{L}$ occurs when the spin momentum $\vec{S}$ is in the opposite direction of $\vec{L}$. Such a simplified treatment is not possible for a magnetic ion with no unevenly-filled degenerate *d*-set, because its orbital momentum is largely quenched and because the direction and the length of the remnant momentum $\delta\vec{L}$ are unknown. This is why use of the perturbation theory is necessary in predicting the preferred spin orientation in such cases, as discussed in the next section.

**2. Effect of spin exchange on spin orientation**

The $Ni^{2+}$ ions of $NiO_6$ trigonal prisms in the $NiIrO_6$ chains of $Sr_3NiIrO_6$, which is isostructural with $Sr_3NiPtO_6$, show the ||z spin arrangement.[1] This is due to the fact that the low-spin $Ir^{4+}$ ($d^5$, S = 1/2) ions possess uniaxial magnetism, and that the spin exchange between adjacent $Ir^{4+}$ and $Ni^{2+}$ ions is strongly antiferromagnetic.[2] The latter requires a collinear arrangement between their spins, and the preference for the ||z spin orientation arising from the degenerate perturbation at the $Ir^{4+}$ ion is much stronger than that for the ⊥z spin orientation arising from the nondegenerate perturbation at the $Ni^{2+}$ ion.[2]

**3. Weak anisotropy of high-spin $Fe^{2+}$ ions in $FeCl_6$ octahedra**

The $FeCl_6$ octahedra containing high-spin $Fe^{2+}$ ($d^6$, S = 2) ions, found in $RbFeCl_3$ [3] and $FeCl_2{\cdot}2H_2O$,[4] exhibit only weakly anisotropic magnetic properties rather than uniaxial magnetism. An electronic factor contributing to this observation would be that when the *p*-orbitals of the ligand Cl is more diffuse than the *d*-orbitals of Fe. Thus, the *d*-states of the $FeCl_6$ octahedron become weakly split, the magnetic orbitals of $FeCl_6$ become dominated by the ligand *p*-orbitals of Cl, and the SOC constant $\lambda$ is not large.

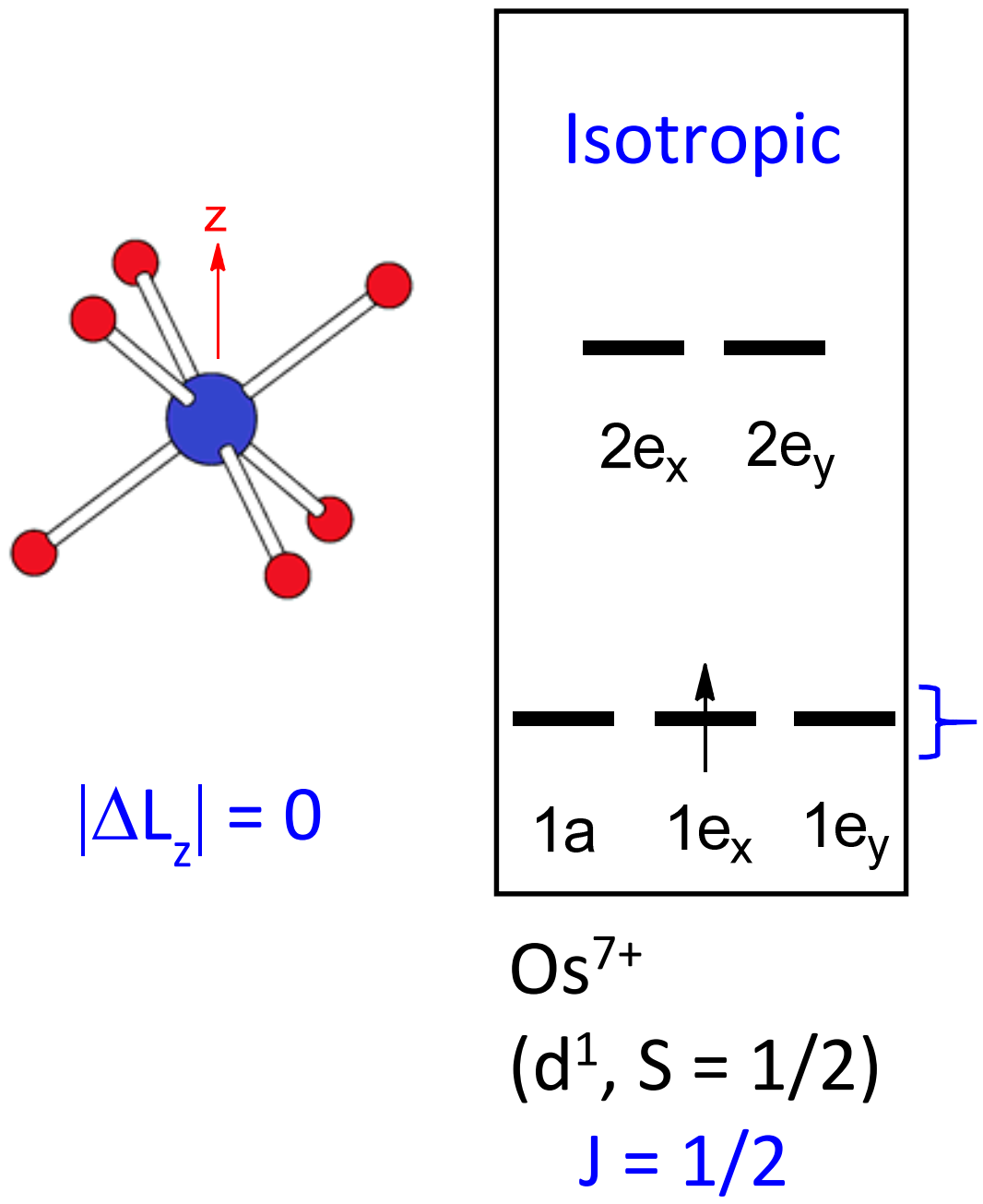

Figure S1.